\documentclass[12pt,a4paper]{article}
\usepackage{amsfonts}
\usepackage{color}
\usepackage{amssymb}
\usepackage{amsmath}
\usepackage{graphicx,color}
\usepackage{amsthm}
\usepackage[latin9]{inputenc}
\usepackage{enumerate}
\usepackage{bm}
\usepackage{fancyhdr}
\usepackage{pdfsync}
\newtheorem*{obs}{Observation}

\newtheorem*{lemma*}{Lema}

\numberwithin{equation}{section}

\newcommand{\xes}{\textrm{\textbf{x}}}
\newcommand{\yes}{\textrm{\textbf{y}}}
\newcommand{\zes}{\textrm{\textbf{z}}}
\newcommand{\tes}{\textrm{\textbf{t}}}

\newcommand{\e}{{\rm e}}

\newcommand{\atan}{{\rm  arctan}}

\newcommand{\Ref}[1]{(\ref{#1})}
\date{}
\begin{document}

\title{\bf Quantum vacuum interaction between two sine-Gordon kinks}

\author{M. Bordag$^1$\footnote{bordag@itp.uni-leipzig.de}, and J. M. Mu$\tilde{\rm n}$oz-Casta$\tilde{\rm n}$eda$^1$\footnote{jose.munoz-castaneda@uni-leipzig.de}\\
\footnotesize{{\sl $^1$Institut für Theoretische Physik, Universität Leipzig, Germany.}}}

\maketitle

\begin{abstract}
We calculate the quantum vacuum interaction energy between two kinks of the sine-Gordon equation. Using the $TGTG$--formula, the problem is reduced to the known formulas for quantum fluctuations in the background of a single kink. This interaction induces an attractive force between the kinks in parallel to the Casimir force between conducting mirrors.
\end{abstract}

\section{Introduction.}
Topological solitons belong to the most interesting topics in quantum field theory as models for extended objects like strings, domain walls and
baryons. The interaction of these objects, especially the
scattering on one another, follows from their properties as
classical fields. In addition, there is an interaction due to the
vacuum fluctuations of quantum fields coupled to the background of
the topological solitons. This is a vacuum quantum effect  in
complete analogy to the Casimir effect between conducting
surfaces. The investigation of this interaction was so far quite difficult due to the inherent ultraviolet divergences present in
the intermediate steps and the complicated procedure of their
removal. The situation changed with the appearance some years ago
of the new T-matrix (or scattering) representation of the vacuum interaction
energy \cite{bulg06-73-025007,emig06-96-080403}. It is also called {\it TGTG-formula} and we follow this notation. It allows to compute the Casimir force between objects of
complicated shape with moderate computational effort. In \cite{kenn08-78-014103}, the method was generalized to background fields that give rise to potentials with compact support. In the present paper we are going to apply this method to the vacuum interaction of two kinks of the sine-Gordon ($SG$) equation.

Since the appearance of the $DHN$ formula in the middle 70's (\cite{Dashen:1974cj,Dashen:1975xh}), where the one loop correction to the mass of the $\lambda\phi^4$ kink was firstly computed, many progress has been made in developing computational methods to calculate 1--loop corrections to the mass of topological solitons. Heat kernel and zeta function techniques (\cite{hawk76-13-191,eliz94b,Vassilevich:2003xt}) have been very powerful to compute one loop corrections to masses of topological defects even when the explicit classical solution for the topological defect is not known. An complete account for the vacuum energy in a continuous background field in terms of scattering data was given in \cite{bord95-28-755}. Remarkable recent results using zeta functional techniques are in the papers by Alonso-Izquierdo, Guilarte {\it et al} \cite{AlonsoIzquierdo:2011dy,AlonsoIzquierdo:2008rk,AlonsoIzquierdo:2003gh,Izquierdo:2007xu}.

Classical interaction between topological objects in field theory is a very well known topic, for references see the book \cite{mantonbook}. As related to the present paper we mention \cite{sutc93-393-211,jack75-12-1643}, where the interaction between two $SG$--kinks was considered; the classical and quantum (in terms of collective coordinate) scattering for instance.

The interaction between classical objects like kinks receives contributions from the vacuum energy of the quantum fluctuations of these fields. For one object, these are the quantum corrections to the mass mentioned above. If one has two objects, their  common mass receives similar corrections. We are interested in that part, which depends on the separation and which is responsible for the force acting between them. Of course, this part can be calculated starting from the complete quantum energy  of the compound object. An example is the vacuum interaction between two delta functions in \cite{Bordag:1992cm}. However,  this makes it necessary to solve the spectral problem in the background of two kinks. This is possible, in principle, but quite cumbersome. Using the $TGTG$--formula, this problem can be solved using the knowledge from the problem for a single object. In this way, the problem becomes much simpler, especially in case the scattering for a single object is known like in the $SG$--model which we consider here.

In the next section we introduce the necessary notation and basic formulas of the $SG$--model. In Section 3 we consider  two kinks and their classical interaction. In Section 3, we first define the
setup for the quantum fluctuations, including the $TGTG$--formula, and than calculate the quantum interaction. Conclusions are given in the last section.

\section{Units and dimensions in the sine-Gordon model}
We are interested in the calculation of the quantum corrections to a classical energy. Hence it is useful to take a system of units in which
\begin{equation*}
c=1,\quad {\rm and}\quad \hbar\neq 1.
\end{equation*}
Therefore $L=T$ and $M\neq L$: $[\hbar]=M\cdot L$. With this system of units, for a $1+1$ dimensional field theory, the dimension of the lagrangian density is $[\overline{\cal L}_{1+1}]=M\cdot L^{-1}$.
  The most general scalar field theory is given by a lagrangian density of the form\footnote{Quantities with units will be marked by an overline.}
\begin{equation*}
  \overline{{\cal L}}(\overline\phi)={1\over 2}\partial_\mu\overline\phi^*\partial^\mu\overline\phi-U(\overline\phi).
\end{equation*}
Hence, a scalar field in a $1+1$ dimensional space-time will have the units $[\overline\phi_{1+1}]=M^{1/2}\cdot L^{1/2}$, and the energy
\begin{equation*}
  E=\int d\overline x\left[ {1\over 2}((\partial_{\overline t}\overline\phi)^2+(\partial_{\overline x}\overline\phi)^2)+U(\overline\phi)\right]
\end{equation*}
has units of mass: $[\overline E]=M$. Following the notation used in \cite{raja82b} the sine-Gordon model is described by the lagrangian density
\begin{equation*}
  \overline{{\cal L}}={1\over2}\partial_\mu\overline{\phi} \partial^\mu\overline{\phi} +\frac{m^4}{\lambda}\left[\cos \left(\frac{\sqrt{\lambda}}{m}\overline{\phi}\right)-1\right].
\end{equation*}
In the system of units selected and taking into account that the sine-Gordon model is a $1+1$ dimensional field theory, the constants appearing in the sine-Gordon model have the following units:
\begin{equation*}
  [\lambda]=M^{-1}\cdot L^{-3};\quad [m]=L^{-1}.
\end{equation*}
To perform numerical calculations is useful to rewrite the sine-Gordon model in terms of pure non-dimensional quantities $\{\phi,\, x,\, t\}$. Again following \cite{raja82b}, we can introduce the non-dimensional quantities using the constants of the problem:
\begin{equation}
  x\equiv m\overline x;\quad t\equiv m\overline t;\quad\phi\equiv\frac{\sqrt{\lambda}}{m}\overline\phi\,.
\end{equation}
Hence, the dimensional lagrangian, can be written as
\begin{equation}
  \overline{\cal L}=\frac{m^4}{\lambda}{\cal L},
\end{equation}
being ${\cal L}$ the dimensionless lagrangian written in terms of non-dimensional quantities:
\begin{equation}
  {\cal L}={1\over 2}(\partial_\mu\phi)(\partial^\mu\phi)+(\cos(\phi)-1).
\end{equation}
The energy functional written in terms of non dimensional quantities, takes the form
\begin{equation}
  E=\frac{m^3}{\lambda}\int dx \left[{1\over2}\left((\partial_t\phi)^2+(\partial_x\phi)^2\right)-(\cos(\phi)-1)\right]\label{energyfunc}
\end{equation}
and the action functional is now
\begin{equation}
  S(\phi)=\frac{m^2}{\lambda}\int dx\,dt{\cal L}.
\end{equation}
The constants of the original lagrangian  allow to define characteristic action and energy scale for the sine-Gordon model,
\begin{equation}
  s_c=m^2/\lambda,\quad\epsilon_c=m^3/\lambda.
\end{equation}
These characteristic scales are used to write the loop expansion for quantum fluctuations around any background field, in terms of dimensionless coefficients. Since the loop expansion is a series in powers of $\hbar$, the loop expansion of the energy of a quantum configuration ($E_Q$), written in terms of dimensionless coefficients and quantities, has the form:
\begin{equation}\label{2.7}
  E_{Q}=\frac{m^3}{\lambda}\sum_{k=0}^{\infty}\left(\frac{\hbar\lambda}{m^2}\right)^k E_k(\phi_Q).
\end{equation}
In this last expression, $E_0$ is the classical energy corresponding to the classical configuration that gives rise to quantum fluctuations, and $E_1$ is the 1--loop correction to the classical energy. At the classical level, the sine-Gordon model has a dimensional coupling constant, but at the quantum level the loop expansion has a dimensionless coupling given by $\hbar\lambda/m^2$, that is assumed to be small in order to give the loop expansion a sense.
Hence we assume that $\hbar\lambda/m^2\ll1$. Our purpose is to compute the distance dependence of 1--loop correction to the force between two kinks in the large separation regime, using  the $TGTG$--formula. We mention, that the same dimensions analysis is valid for the $\lambda\phi^4$ since it is just a truncation of the cosine series for the potential in the sine-Gordon model.

\section{Kink configurations in the classical sine-Gordon model}
Following notation used in \cite{raja82b}, the dimensionless action for the sine-Gordon model is given by
\begin{equation}
S\left(\phi\right)=\int dx\,dt\left({1\over 2}(\partial_\mu\phi)(\partial^\mu\phi)+(\cos(\phi)-1)\right).
\end{equation}
The equation of motion arising from the preceding action functional is the very well known sine-Gordon equation
\begin{equation}
  \partial^2\phi+\sin\left(\phi\right)=0.
\end{equation}
From this equation, we obtain the kink and anti-kink solutions:
\begin{eqnarray}
  \phi_K(x)&=&4\atan\left(e^{x}\right)\\
  \phi_{\overline{K}}(x)&=&-4\atan\left(e^{x}\right)=-\phi_K(x).
\end{eqnarray}
The energy functional for the sine-Gordon model is given by equation (\ref{energyfunc}). For static field configurations the dimensionless energy functional can be re-written as:
\begin{equation}
  E[\phi]=\int dx \left[{1\over2}(\partial_x\phi)^2+2\sin^2\left(\frac{\phi}{2}\right)\right]={1\over 2}\int dx \left(\partial_x\phi-2\sin\left(\frac{\phi}{2}\right)\right) -4\left[\cos\left(\frac{\phi}{2}\right)\right]_{-\infty}^\infty \label{energystat}
\end{equation}
Last equality is the Bogomolnyi expression for the energy functional, and gives rise to the first order equations for stable classical solutions with finite energy
\begin{equation}
  \partial_x\phi-2\sin\left(\frac{\phi}{2}\right)=0.
\end{equation}
Using the explicit expression for the energy functional, it is very easy to compute its dimensionless numerical value for a static field configuration given by addition of two kinks with different centers of mass separated a distance $a$:
\begin{equation}\label{3.7}
  \phi_{2K}(x;a)=\phi_K(x+a/2)+\phi_K(x-a/2).
\end{equation}
Calling $\phi_1(x)=\phi_K(x+a/2)$ and $\phi_2(x)=\phi_K(x-a/2)$, and knowing that from the Bogmolny first order equation $\partial_x\phi_K=\sin(\phi_K/2)$ if we take into account that $E[\phi_1]=E[\phi_2]=E_K$ (where $E_K$ is the energy of one kink which is independent of the position of the center of mass), the classical energy for the configuration $\phi_{2K}$ is given by:
\begin{equation}
  E[\phi_1+\phi_2]=2E_K+4\int dx\sin\left(\frac{\phi_1}{2}\right)\sin\left(\frac{\phi_2}{2}\right)\left(1+\cos\left(\frac{\phi_1+\phi_2}{2}\right)\right).\label{eclas2k}
\end{equation}
Only the last term in the right hand side of (\ref{eclas2k}) depends on the distance between kinks. Therefore the last term on the right hand side of (\ref{eclas2k}) determines the classical interaction energy between two kinks $\Delta_cE_{{\rm int}}(\phi_1,\phi_2)$:
\begin{equation}
  \Delta_cE_{{\rm int}}(\phi_1,\phi_2)=4\int dx\sin\left(\frac{\phi_1}{2}\right)\sin\left(\frac{\phi_2}{2}\right)\left(1+\cos\left(\frac{\phi_1+\phi_2}{2}\right)\right).
\end{equation}
Using the explicit expression for the kink solution given above and simplifying, we obtain the classical interaction energy density between two kinks
\begin{equation}
  \Delta_c\epsilon_{{\rm int}}(K,K)=8\frac{\sinh^2(x)}{\cosh^2(x+a/2)\cosh^2(x-/2)}.
\end{equation}
Integrating the energy density we obtain the classical interaction energy
\begin{equation}\label{3.11}
  \Delta_cE_{{\rm int}}(K,K)=16\frac{\sinh(a)-a}{\sinh(a)\left(\cosh(a)-1\right)}.
\end{equation}
The same calculation can be performed for the interaction between a kink and an anti-kink (note that $E_{\overline{K}}=E_K$)  just replacing $\phi_K(x-a/2)$ by $-\phi_K(x-a/2)=\phi_{\overline{K}}(x-a/2)$ giving rise to the classical interaction energy density
\begin{equation}
  \Delta_c\epsilon_{{\rm int}}(K,\overline{K})=-8\frac{\cosh^2(x)}{\cosh^2(x+a/2)\cosh^2(x-/2)}.
\end{equation}
After integrating the energy density, the classical interaction energy between a kink and an anti-kink is obtained:
\begin{equation}
  \Delta_c E_{{\rm int}}(K,\overline{K})=-16\frac{\sinh(a)+a}{\sinh(a)\left(\cosh(a)+1\right)}
\end{equation}
\begin{figure}[h!]\label{fig3}
\centerline{\includegraphics[height=5cm]{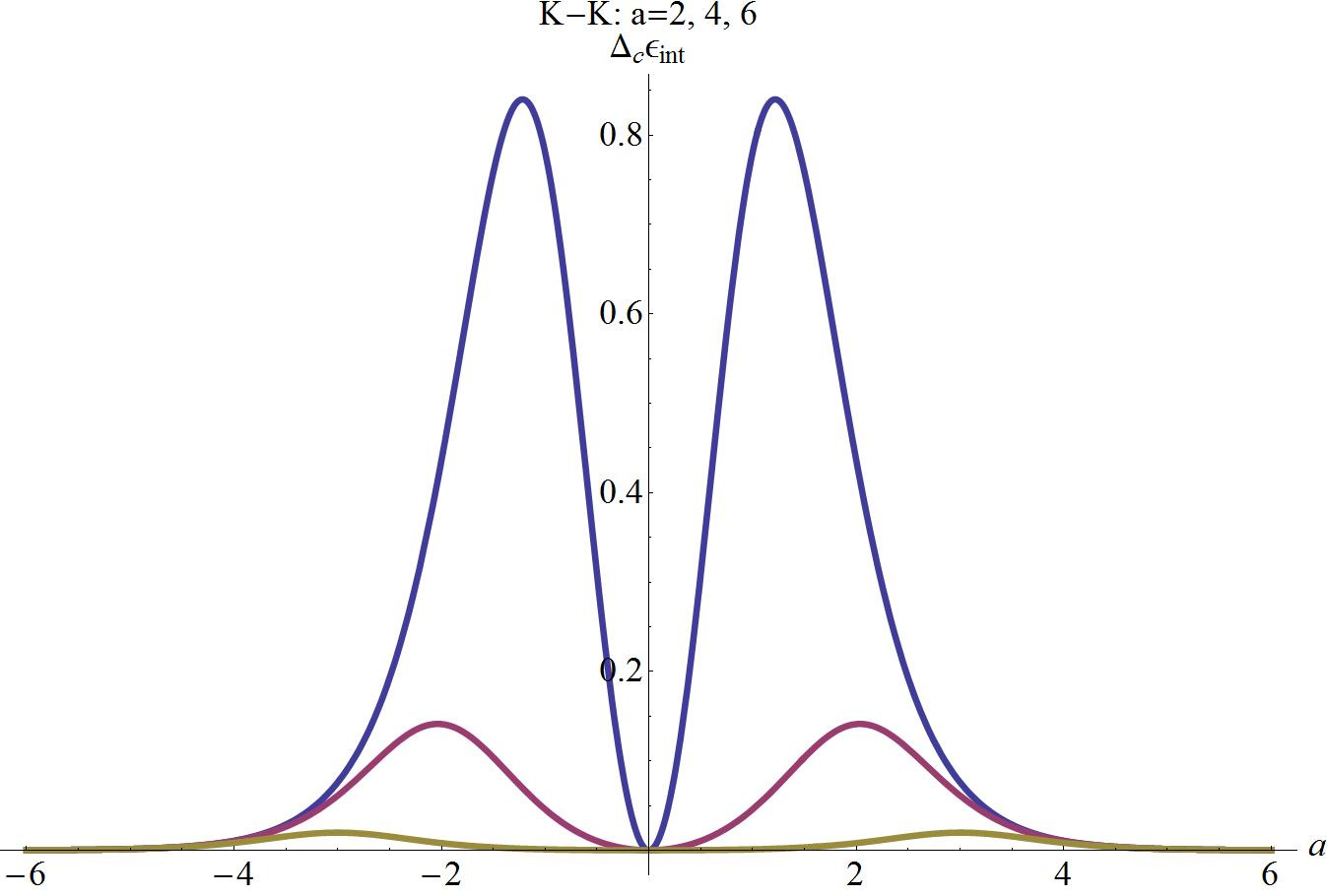}\includegraphics[height=5cm]{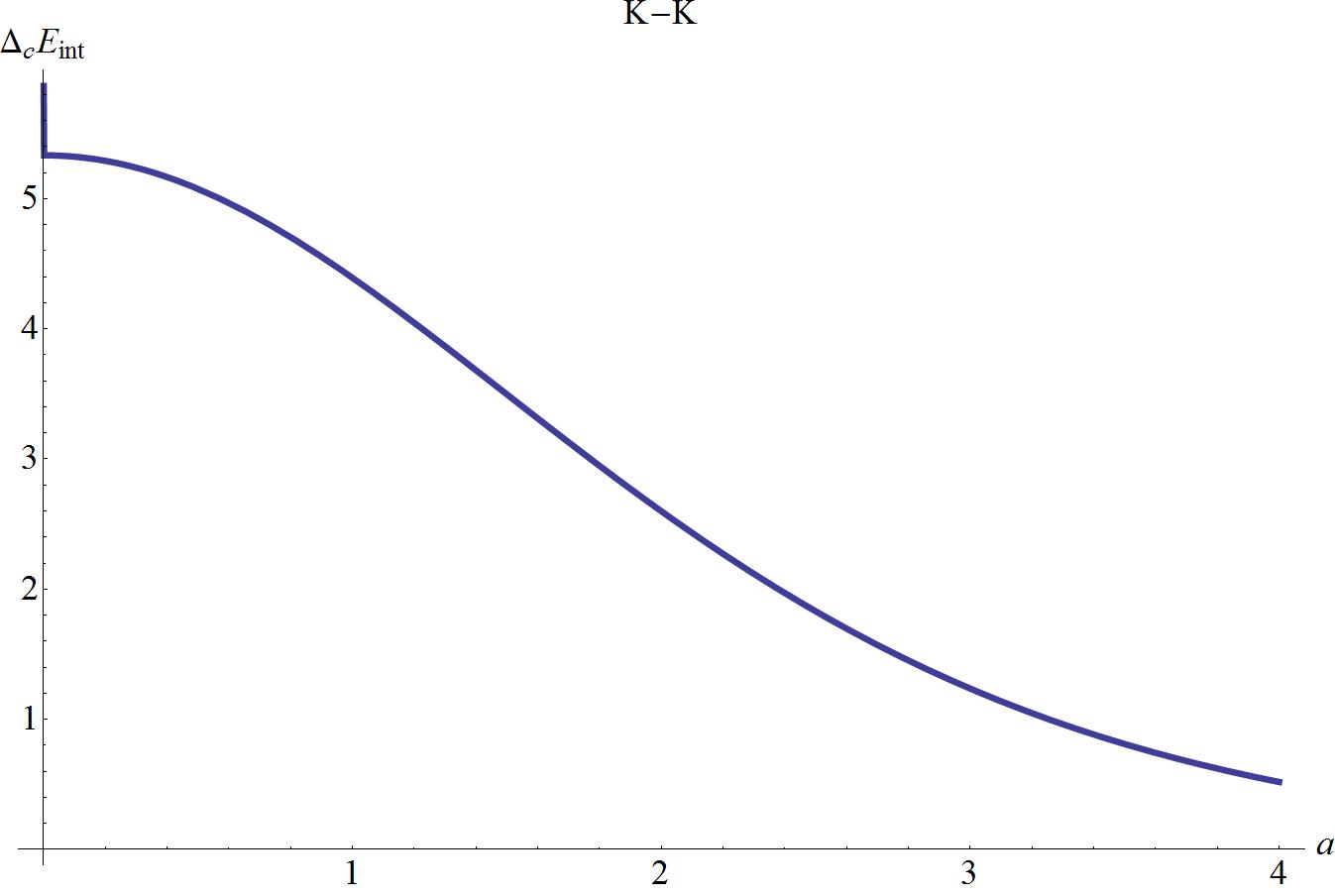}}
\caption{\footnotesize{Classical interaction energy density (left) and classical interaction energy (right) for two kinks.}}
\end{figure}
\begin{figure}[h!]
\centerline{\includegraphics[height=5cm]{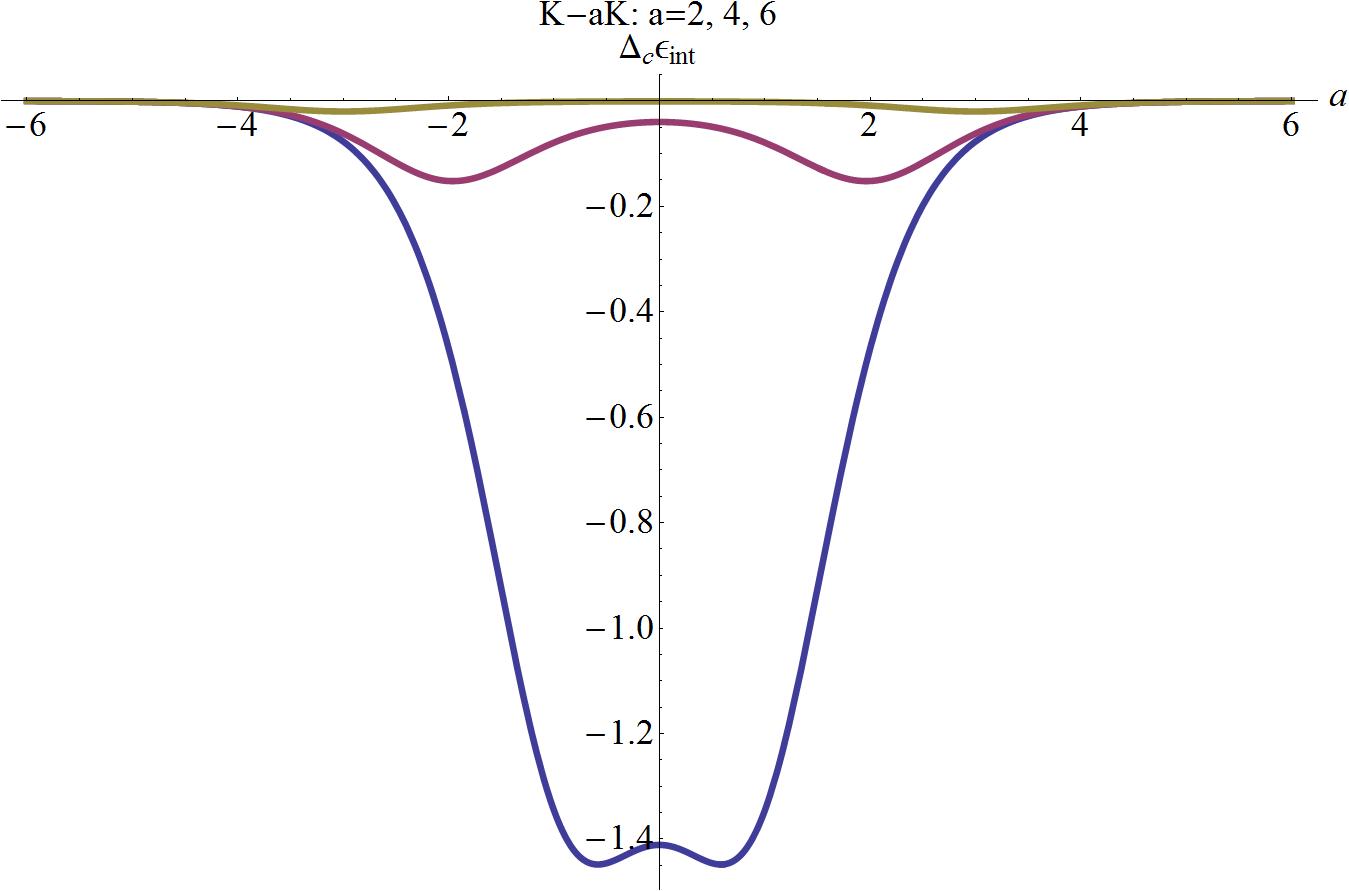}
 \includegraphics[height=4.9cm]{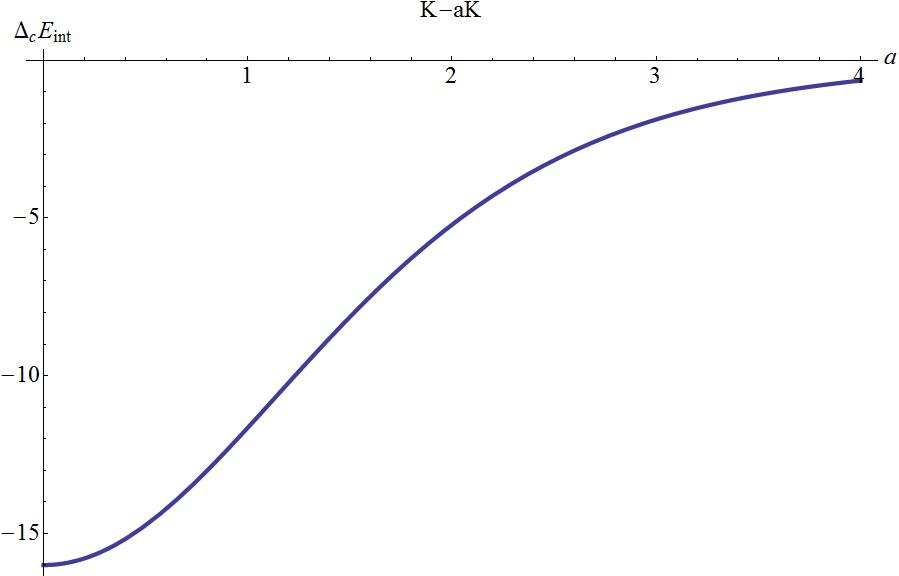}}
\caption{\footnotesize{Classical interaction energy density (left) and classical interaction energy (right) for a kink and an anti-kink.}}
\end{figure}

We mention that Eq. \ref{3.7} is not the only way to introduce two kinks which are not solutions of the equation of motion. Another possibility, aimed from the exact solution with two moving kinks, was used in \cite{sutc93-393-211}, Eq. (3.1),
\begin{equation}\label{sutc}
    \phi_{2K}=4\arctan\left(e^{x-\frac{a}{2}}-e^{-x-\frac{a}{2}}\right),
\end{equation}
which has, however, a larger classical energy than the configuration \ref{3.7} considered in this paper.

\section{Quantum vacuum interaction of two kinks}
In the $TGTG$--method for calculating the vacuum interaction energy it is assumed the background potential to be a sum of two\footnote{The original $TGTG$--method was developed to compute the vacuum interaction energy between dielectric bodies of arbitrary shape. In this picture each if those objects were represented by a classical potential with compact support.},
\begin{equation}\label{V12}
   V(\xes)=V_1(\xes)+V_2(\xes).
\end{equation}
Originally, it was assumed that both parts of the potential have non--intersecting compact supports. In fact, this restriction can be released. In that case one cannot reduce the problem to the scattering data but since we have explicit formulas for the scattering at any separation, this  restriction is not effective in our case. In the original papers the method was used in  (3+1) dimensions. So we will give the corresponding formulas for (1+1) dimension below. These are easier, but specific since we have a scattering problem on the whole axis in place of the half-axis of a radial variable.

\subsection{Vacuum interaction between extended objects in $1+1$ dimensional field theories.}
A scalar field theory over the real line is described by the dimensionless action functional
\begin{equation*}
  S(\phi)=\int d^2x\left({1\over 2}\partial_\mu\phi\partial^\mu\phi-U(\phi)\right).
\end{equation*}
If $\Phi_0$ is a classical field configuration the formula for the vacuum energy of the small quantum oscillations around this classical background is given by
 \begin{equation}
  E_0= \frac{i}{2}\int_{0}^\infty\frac{d\omega}{\pi}{\rm Tr}\ln\left({\cal G}_\omega^{(V)}\right)
\end{equation}
(see, e.g., Eq. (3.112) in \cite{BKMM}), where $V(x)$ is the effective potential defined by the background classical field
\begin{equation}
  V(x)=\left.\frac{d^2 U}{d\phi^2}\right|_{\Phi_0},
\end{equation}
and ${\cal G}_\omega^{(V)}$ is the Green functional associated to the Schr\"odinger problem defined by the effective potential for the quantum fluctuations around the classical background. The euclidean formula for the vacuum energy is given by:
\begin{equation}
  E_0=-\frac{1}{2}\int_{0}^\infty\frac{d\xi}{\pi}{\rm Tr}\ln\left({\cal G}_{i\xi}^{(V)}\right)
\end{equation}

\paragraph{Remark on the euclidean formulation.} When the quantum mechanical system for one particle states arising from the quantum field theory has bound states we must be careful in all the manipulations that involve free field operators. All calculations must be done in the euclidean rotated system in order to avoid pathological expressions in which operators acting on different Hilbert spaces are multiplied (these multiplications are not defined). When euclidean rotation is taken all the spectra are continuous and the only restriction we must take into account is the possibility of a lower bound for the imaginary frequency $\xi$.
\par
In order to use just a potential that goes to 0 when $x\rightarrow\pm\infty$ we define the scattering potential $\widetilde{V}(x)$ in terms of $m^2\equiv\lim_{x\rightarrow\pm\infty}V(x)$ as
\begin{equation}
\widetilde{V}(x)\equiv V(x)-m^2.
\end{equation}
The euclidean Schr\"odinger problem associated with the one particle states of the field theory is described by the differential equation
\begin{equation}
  \left(\xi^2-\frac{d^2}{dx^2}+\widetilde{V}(x)+m^2\right)\phi_\xi(x)=0\label{qmsys},
\end{equation}
and the kernel $G_{i\xi}^{(V)}(x,y)$ of ${\cal G}_{i\xi}^{(V)}$ is given\footnote{In order to simplify the notation, from now on and unless it induces   errors, we will denote the euclidean rotated operators kernels and functions just with the subindex $\xi$ instead of the subindex $i\xi$. We will use subindex $\omega$ to denote the real frequencies, and hence will refer to the standard theory (non Wick rotated) instead of its euclidean version.} by the equation
\begin{equation}
  \left(\xi^2-\frac{d^2}{dx^2}+\widetilde{V}(x)+m^2\right)G_{\xi}^{(V)}(x,y)=\delta(x-y).
\end{equation}
\begin{obs}
  Note that the Green function for the potentials $V(x)$ and $\widetilde{V}(x)$ are related by the equality
\begin{equation}\label{4.6}
  G_{\xi}^{(V)}(x,y)=G_{\overline\xi}^{(\widetilde{V})}(x,y),\quad\overline\xi=\sqrt{\xi^2+m^2}.
\end{equation}
Taking this into account from now on we will only manipulate Green function associated with the scattering potential $\widetilde{V}(x)$ associated to $V(x)$, and keep in mind that the vacuum energy for the background defined by $V(x)$ can be computed in terms of the Green function associated with $\widetilde{V}(x)$ changing the integration interval in $\xi$:
\begin{equation}
  E_0[V]=-\frac{1}{2}\int_{m}^\infty\frac{d\xi}{\pi}\frac{\xi}{\sqrt{\xi^2-m^2}}{\rm Tr}\ln\left({\cal G}_{\xi}^{(\widetilde{V})}\right).\label{evacvm}
\end{equation}
\end{obs}
To obtain an expression for the kernel $G_{\xi}^{(\widetilde{V})}(x,y)$ it is useful to go back to the non-euclidean theory and study the standard scattering problem. If $u_k(x)$ and $v_k(x)$ are two independent solutions of the problem defined by
\begin{equation}
  \left(-\frac{d^2}{dx^2}+\widetilde{V}(x)\right)\psi(x)=k^2\psi(x),
\end{equation}
then the kernel ($\omega^2=k^2$ for the scattering potential $\widetilde{V}$) $G_\omega^{(\widetilde{V})}(x,y)$ of ${\cal G}_\omega^{(\widetilde{V})}$ is given by
\begin{equation}
  G^{(\widetilde{V})}_\omega(x,x')=-\frac{u_k(x_<)v_k(x_>)}{W[u_k,v_k]},
\end{equation}
being $W[u_k,v_k]=u_k(x)v_k'(x)-u_k'(x)v_k(x)$ the Wronskian of the two solutions\footnote{Observe that since $u_k(x)$ and $v_k(x)$ are independent solutions with the same eigenvalue their Wronskian is constant and non zero \cite{galindopascual}}. From scattering theory, if $k$ is in the continuum spectrum, then $u_k(x)$ and $v_k(x)$ correspond to the right and left handed scattering solutions, whose asymptotic behavior is given by
\begin{equation}
  v_k(x)=\left\{
        \begin{array}{ll}
          \e^{ikx}+r_r\e^{-ikx}, & x\rightarrow-\infty, \\
          t_r\e^{ikx}, & x\rightarrow\infty,
        \end{array}
      \right.
\end{equation}
and
\begin{equation}
  u_k(x)=\left\{
        \begin{array}{ll}
          t_l\e^{-ikx}, & x\rightarrow-\infty, \\
          r_l\e^{ikx}+\e^{-ikx}, & x\rightarrow\infty.
        \end{array}
      \right.
\end{equation}
Using the Lippmann-Schwinger equations we can write the differential equation for the propagator $G_\xi^{(\widetilde{V})}(x,x')$ in an integral form, in terms of the potential $\widetilde{V}(x)$ and the free Green function $G_\xi^{(0)}(x,x')$:
\begin{equation}
  G_\xi^{(\widetilde{V})}(x,x')=G_\xi^{(0)}(x,x')-\int dx_1 G_\xi^{(0)}(x,x_1)\widetilde{V}(x_1)G_\xi^{(\widetilde{V})}(x_1,x').
\end{equation}
Defining the $T$ operator for the potential $\widetilde{V}$ in coordinates as
\begin{equation}
  G_\xi^{(\widetilde{V})}(x,x')=G_\xi^{(0)}(x,x')-\int dx_1dx_2
  G_\xi^{(0)}(x,x_1)T^{(\widetilde{V})}_\xi(x_1,x_2)G_\xi^{(0)}(x_2,x'),
\end{equation}
one deduces that
\begin{equation}
  \widetilde{V}(x)G_\xi^{(\widetilde{V})}(x,x')=\int dx_1 T^{(\widetilde{V})}_\xi(x,x_1) G_\xi^{(0)}(x_1,x'),\label{vgtg}
\end{equation}
and, in operator notation,
\begin{equation*}
  {\cal G}_\xi^{(\widetilde{V})}={\cal G}_\xi^{(0)}-{\cal G}_\xi^{(0)}\cdot{\cal T}_\xi\cdot{\cal G}_\xi^{(0)}=
  {\cal G}_\xi^{(0)}-{\cal G}_\xi^{(0)}\cdot\widetilde{{\cal V}}\cdot{\cal G}_\xi^{(\widetilde{V})},
\end{equation*}
\begin{equation}
  \Rightarrow {\cal G}_\xi^{(\widetilde{V})}=\frac{1}{1+{\cal G}_\xi^{(0)}\cdot\widetilde{{\cal V}}}\cdot{\cal G}_\xi^{(0)}=
  {\cal G}_\xi^{(0)}\cdot\frac{1}{1+\widetilde{{\cal V}}\cdot{\cal G}_\xi^{(0)}}.
\end{equation}
Suppose now that $\widetilde{V}(x)$ is split into two parts, $\widetilde{V}(x)=\widetilde{V}_1(x)+\widetilde{V}_2(x)$. In the next section we are going to apply this splitting to the potential of two kinks,

\begin{equation}\label{2k}
    V(x)=V_1(x)+V_2(x)=m^2+\widetilde{V}_1(x)+\widetilde{V}_2(x),
\end{equation}
where now $m^2=m_1^2+m_2^2$ being $m_i^2\equiv\lim_{x\rightarrow\infty}V_i(x)$. Using the preceding formula, we can easily write
\begin{equation}
  {\cal G}_\xi^{(\widetilde{V}_1+\widetilde{V}_2)}=\frac{1}{1+ {\cal G}_\xi^{(0)}\cdot\widetilde{{\cal V}}_1}\cdot\frac{1}{1-\frac{1}{1+ {\cal
  G}_\xi^{(0)}\cdot\widetilde{{\cal V}}_1}{\cal G}_\xi^{(0)}\cdot\widetilde{{\cal V}}_1\frac{1}{1+ {\cal G}_\xi^{(0)}\cdot\widetilde{{\cal
  V}}_2}{\cal G}_\xi^{(0)}\cdot\widetilde{{\cal V}}_2}\cdot\frac{1}{1+ {\cal G}_\xi^{(0)}\cdot\widetilde{{\cal V}}_2}\cdot {\cal
  G}_\xi^{(0)}.\label{gtotal}
\end{equation}
Introducing expression (\ref{gtotal}) into equation (\ref{evacvm}) we obtain an expression for vacuum energy of a classical background given by a potential $V(x)$ that splits into two parts $V(x)=V_1(x)+V_2(x)$,
\begin{eqnarray}
  E_0&=&\nonumber{-\frac{1}{2}\int_{m}^\infty\frac{d\xi}{\pi}\frac{\xi}{\sqrt{\xi^2-m^2}}\left({\rm Tr}\ln\left(\frac{{\cal
  G}_\xi^{(0)}}{{1+\cal G}_\xi^{(0)}\widetilde{{\cal V}}_1}\right)+{\rm Tr}\ln\left(\frac{{\cal G}_\xi^{(0)}}{{1+\cal
  G}_\xi^{(0)}\widetilde{{\cal V}}_2}\right)\right.}\\
  &&- {\rm Tr}\ln\left({\cal G}_\omega^{(0)}\right)-{\rm Tr}\ln\left(1-{\cal M}_\xi\right)\Bigg),\label{totvac}
\end{eqnarray}
where the operator ${\cal M}_\xi$ is defined as
\begin{equation}
  {\cal M}_\xi={\cal N}^{(1)}_\xi\cdot {\cal N}^{(2)}_\xi;\quad {\cal N}^{(i)}_\xi= \frac{1}{{1+\cal G}_\omega^{(0)}\cdot\widetilde{{\cal V}}_i}\cdot{\cal G}_\omega^{(0)}\widetilde{{\cal V}}_i.
\end{equation}
Looking at equation (\ref{totvac}) one notices that  the first three terms do not depend on the distance between the objects that potentials $V_1(x)$ and $V_2(x)$ represent. Hence, they will not contribute to the Casimir force between them, i. e., they do not enter in the vacuum interaction between the objects represented by $V_1(x)$ and $V_2(x)$. Therefore the last term of equation (\ref{totvac}) gives the vacuum interaction energy between the objects represented by $V_1(x)$ and $V_2(x)$,
\begin{equation}\label{4.23}
  E_{{\rm int}}^{(0)}=\frac{1}{2}\int_{m}^\infty\frac{d\xi}{\pi}\frac{\xi}{\sqrt{\xi^2-m^2}}{\rm Tr}\ln\left(1-{\cal M}_\xi\right).
\end{equation}
Knowing that in operator notation the equation (\ref{vgtg}) takes the form $\widetilde{{\cal V}}\cdot{\cal G}_\xi^{(\widetilde{V})} ={\cal T}_\xi^{(\widetilde{V})}\cdot{\cal G}_\xi^{(0)}$, we can rewrite the operator ${\cal M}_\xi$ just in terms of the two ${\cal T}$-operators (one for each potential), and the vacuum Green functions using the relation
\begin{equation}
  {\cal N}^{(i)}_\xi= \frac{1}{{1+\cal G}_\xi^{(0)}\cdot\widetilde{{\cal V}}_i}\cdot{\cal G}_\omega^{(0)}\cdot\widetilde{{\cal V}}_i = {\cal
  G}_\xi^{(\widetilde{V}_i)}\cdot\widetilde{{\cal V}}_i={\cal G}_\xi^{(0)}\cdot{\cal T}_\xi^{(i)}
\end{equation}
for the ${\cal N}^{(i)}_\xi$ operators.

The operator notation has been very useful to obtain a general expression for the vacuum interaction energy. However, in order to explicitly compute the quantum vacuum interaction energy between two extended objects in one dimensional quantum field theory, we must give an expression of the energy in terms of the kernels for the corresponding operators. As usual, we will denote the kernel of an operator ${\cal K}$ by the corresponding capital letter $K(x,x')$. Hence,
\begin{equation}
  M_\xi(x,x')=\int dx'' N_\xi^{(1)}(x,x'')N_\xi^{(2)}(x'',x')\label{mker},
\end{equation}
and
\begin{equation}
  N_\xi^{(i)}(x,x')=\int dx'' G_\xi^{(0)}(x,x'')T_\xi^{(i)}(x'', x')\label{nker}.
\end{equation}
Since we are in the perturbative regime of the field theory we can assume that $||{\cal M}_\xi||< 1$, and hence write\footnote{The condition $||{\cal M}_\xi||< 1$ is closely related to the existence of bound states in the ordinary quantum mechanical system for the on-particle states od the corresponding quantum field theory.} $\ln({\bf 1}-{\cal M}_\xi)\simeq-{\cal M}_\xi-{\cal M}_\xi^2/2+ {\cal O}\left(||{\cal M}_\xi||^3\right)$. Under this assumption we can write the interaction energy as
\begin{eqnarray}
  E_{{\rm int}}^{(0)}&\simeq&\nonumber{-\frac{1}{2}\int_{m}^\infty\frac{d\xi}{\pi}\frac{\xi}{\sqrt{\xi^2-m^2}}{\rm Tr}\left({\cal M}_\xi\right)-\frac{1}{4}\int_{m}^\infty\frac{d\xi}{\pi}\frac{\xi}{\sqrt{\xi^2-m^2}}{\rm Tr}\left({\cal
  M}_\xi^2\right)}\\
  &=&\nonumber{-\frac{1}{2}\int_m^\infty\frac{d\xi}{\pi}\frac{\xi}{\sqrt{\xi^2-m^2}}\int dx
  M_\xi(x,x)}\\
  &&-\frac{1}{4}\int_m^\infty\frac{d\xi}{\pi}\frac{\xi}{\sqrt{\xi^2-m^2}}\int dxdx' M_\xi(x,x')M_\xi(x',x)+ {\cal O}\left(||{\cal M}_\xi||^3\right).
\end{eqnarray}
Calling
\begin{equation}
E^{{\rm int}}_n=-\int_m^\infty \frac{d\xi}{\pi n}\frac{\xi}{\sqrt{\xi^2-m^2}}{\rm Tr}\left({\cal M}_\xi^n\right),
\end{equation}
we can write the whole series expansion in powers of ${\cal M}_\xi$ as $E_{{\rm int}}^{(0)}=\sum_{n=1}E^{{\rm int}}_n/2$. Using expressions (\ref{mker}) and (\ref{nker}), the general term of the series expansion in power of ${\cal M}_\xi$ can be written as
\begin{eqnarray}
  E^{{\rm int}}_n&=&\nonumber{-\frac{1}{n}\int_m^\infty\frac{d\xi}{\pi}\frac{\xi}{\sqrt{\xi^2-m^2}}\int dx d\xes_n d \yes_n d\tes_n
  d\zes_n\delta(x_1-x)\delta(x_{n+1}-x)}\\
  &\times&\prod_{k=1}^nG^{(0)}_\xi(x_k,y_k)T_\xi^{(1)}(y_k,z_k)
  G^{(0)}_\xi(z_k,t_k)T_\xi^{(2)}(z_k,x_{k+1}),
\end{eqnarray}
being $d\xes_n=dx_1...dx_{n+1}$ (equivalent for $d\yes_n$ and $d\zes_n$). Therefore with this series expansion the loop expansion for the interaction energy up to first order in $\hbar\lambda/m^2$ is written as
\begin{equation}
  E_{{\rm int}}=-\frac{m^3}{\lambda}\left[\Delta_cE_{{\rm int}}+\frac{\hbar\lambda}{2m^2}\left(E^{{\rm int}}_1+E^{{\rm int}}_2+{\cal O}\left(\|{\cal M}_\xi\|^3\right)\right)+{\cal O}\left(\frac{\hbar\lambda}{m^2}\right)\right].\label{4.30}
\end{equation}
For our purpose,  the case in which the potentials $V_1$ and $V_2$ are the same potential with a given displacement, i. e. $V_2(x)=V_1(x-a)$, is of special interest. In this case we have
\begin{equation}
  N_\xi^{(2)}(x,x')=N_\xi^{(1)}(x-a,x'-a)
\end{equation}
and taking into account that for the free field propagator $G_\xi^{(0)}(x,x')=G_\xi^{(0)}(x-x')$ holds, we can immediately write the two first terms of the vacuum interaction energy,
\begin{equation}
  E^{{\rm int}}_1=-\int_m^\infty\frac{d\xi}{\pi}\frac{\xi}{\sqrt{\xi^2-m^2}}\int dx dyN_\xi^{(1)}(x,y)N_\xi^{(1)}(y-a,x-a) \label{e1sto},
\end{equation}
\begin{eqnarray}
  E^{{\rm int}}_2&=&\nonumber{-\frac{1}{2}\int_m^\infty\frac{d\xi}{\pi}\frac{\xi}{\sqrt{\xi^2-m^2}}\int dx dy_1dy_2dy_3N_\xi^{(1)}(x,y_1)N_\xi^{(1)}(y_1-a,y_2-a)}\\
  &\times&N_\xi^{(1)}(y_2,y_3)N_\xi^{(1)}(y_3-a,x-a)\label{e2ndo},
\end{eqnarray}
where now
\begin{equation}
  N_\xi^{(1)}(x,x')=\int dy G_\xi^{(0)}(x-y)T_\xi^{(1)}(y,x')=G_\xi^{V_1}(x,x')V_1(x)\label{nprop}.
\end{equation}
Hence, using the last equality in expression (\ref{nprop}), we can rewrite each term of the vacuum interaction energy in terms also of the propagator for the potential $V_1(x)$. Also, it should be mentioned at this point, that these expressions allow to perform numerical computations in a very friendly way because calculation of vacuum energy with this method requires only performing a numerical integration.

\subsection{$TGTG$--calculation for the vacuum energy between two kinks: first order.}
In order to be able to obtain an expression for the quantum vacuum interaction energy between two kinks or a kink and an anti-kink, we must firstly compute the propagator. The potential that governs the dynamical behavior (up to second order) of the small fluctuations around the kink and anti-kink configuration\footnote{The spectrum of small quantum fluctuations around the kink and the ant-kink configuration is governed by the same potential: the P\"oschl-Teller transparent potential. This means that at the one-loop level both configurations have the same quantum properties. Therefore, one-loop correction for the mass of the kink and the anti-kink are the same, as well as the quantum vacuum interaction between kink and the anti-kink, and between two kinks.}, is the well known transparent P\"oschl-Teller potential,
\begin{equation}
  V(x)=1-\frac{2}{\cosh^2(x)}\,\Rightarrow\,\, \widetilde{V}(x)=-\frac{2}{\cosh^2(x)}.
\end{equation}
The lower bound in the integrations over the euclidean frequency in this particular case will be given by $m=\sqrt{2}$ since each P\"oschl-Teller potential gives a contribution of 1 to $m^2$. This potential has been deeply studied in many areas of physics, and the Schrödinger problem arising for the corresponding scattering potential
\begin{equation}
  \left(-\frac{d^2}{dx^2}-\frac{2}{\cosh^2(x)}\right)\psi_k=\lambda_k\psi_k
\end{equation}
is perfectly known and solved. The corresponding eigenfunctions and spectrum are given by
\begin{eqnarray}
  \psi_{\pm k}&=&\e^{\pm ikx}\left(\tanh(x)\mp ik\right),\quad\lambda_k=k^2\quad{\rm (scattering\,\,states)},\\
  \psi_{\pm i}&=&\e^{\mp x}\left(\tanh(x)\pm k\right),\quad\lambda_i=-1\quad{\rm (semi-bounded\,\,state)}.
\end{eqnarray}
The corresponding Wronskian of two independent scattering solutions with the same eigenvalue is given by
\begin{equation}
  W(\psi_k,\psi_{-k})=\psi_k\psi'_{-k}-\psi'_k,\psi_{-k}=-2ik(1+k^2)=W(k).
\end{equation}
For $k^2=0$, and $k^2=-1$, there is only one independent solution. For the other cases we have two independent solutions for a given eigenvalue $k^2$. The Fourier projection of the Green function in the time coordinate  is given by
\begin{equation}
  G_\omega^{(\widetilde{V})}(x,y)=\frac{1}{ W(k)}\left(\theta(x-y) \psi_k(x)
  \psi_{-k}(y)+ \theta(y-x) \psi_k(y) \psi_{-k}(x)\right),
\end{equation}
where $\theta(x)$ is the Heaviside step function. From the general expression using the eigenfunctions written above, we obtain the expression for the Green function of the P\"oschl-Teller potential,
\begin{equation}
  G_\omega^{(\widetilde{V})}(x,y)=-\frac{e^{ik|x-y|}}{2ik(k^2+1)}\left(\tanh(x)\tanh(y)+ik\left|\tanh(x)-\tanh(y)\right|+k^2\right)
  \label{ptprop},
\end{equation}
and its euclidean version,
\begin{equation}
  G_{\xi}^{(\widetilde{V})}(x,y)=-\frac{e^{-\kappa|x-y|}}{2\kappa(\kappa^2-1)}\left(\tanh(x)\tanh(y)-\kappa\left|\tanh(x)
  -\tanh(y)\right|-\kappa^2\right), \label{eptprop}
\end{equation}
being\footnote{Note that the expressions of the Green functions for the P\"oschl-Teller potential with mass term ($V(x)=1-2\cosh^{-2}(x)$) will be exactly the same as the ones we have in expressions (\ref{ptprop}) and (\ref{eptprop}) in terms of $k$. The difference would be in the dispersion relation: for the case of the P\"oschl-Teller potential with mass term, the dispersion relation is given by $\omega^2=k^2+1$ for the non euclidean case, and $\xi^2=\kappa^2-1$ for the euclidean case.} $\kappa^2=\xi^2$. Introducing the expression (\ref{eptprop}) in the expression for $N_1$ given in the last equality of equation (\ref{nprop}), we obtain from equations (\ref{e1sto}), and (\ref{e2ndo}) two integral expressions for the first two contributions to the vacuum energy that can be computed numerically. The kernel of the operator ${\cal N}_{\xi}$ for a single kink is obtained from the last equality in equation (\ref{nprop}),
\begin{equation}
  N^{(K)}_{\xi}(x,y)=-\frac{\text{sech}^2(x) e^{-\kappa
   |x-y|} \left(-\kappa  |\tanh
   (x)-\tanh (y)|+\tanh (x) \tanh
   (y)-\kappa ^2\right)}{\kappa
   \left(\kappa ^2-1\right)}.\label{nkink}
\end{equation}
Using the expression for $N^{(K)}_\xi$ given in the preceding equation, it is very easy to compute numerically the first two contributions to the vacuum energy by using formulas (\ref{e1sto}) and (\ref{e2ndo}). In order to obtain the correct result we must take into account that the asymptotic value of the P\"oschl-Teller potential is $1$ so the integration interval over $\xi$ is $(\sqrt{2},\infty)$. After numerical calculations we obtain a behavior showed by figures \ref{fig4}. As can be seen, the contribution of $E^{{\rm int}}_2$ is much smaller than $E^{{\rm int}}_1$. This fact means that in this case the series expansion in powers of operator ${\cal M}$ is convergent, and hence $||{\cal M}_\xi||<1$ justifying the expansion  of the logarithm.

\begin{figure}[h!]
\centerline{\includegraphics[height=5cm]{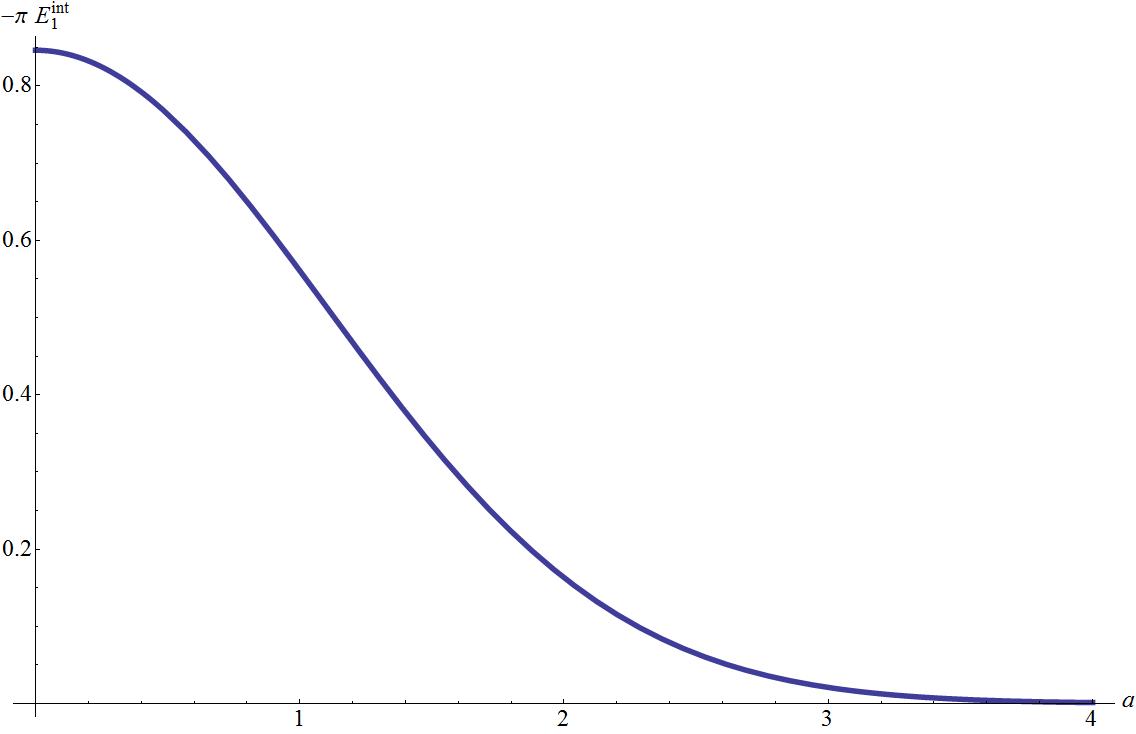}\includegraphics[height=5cm]{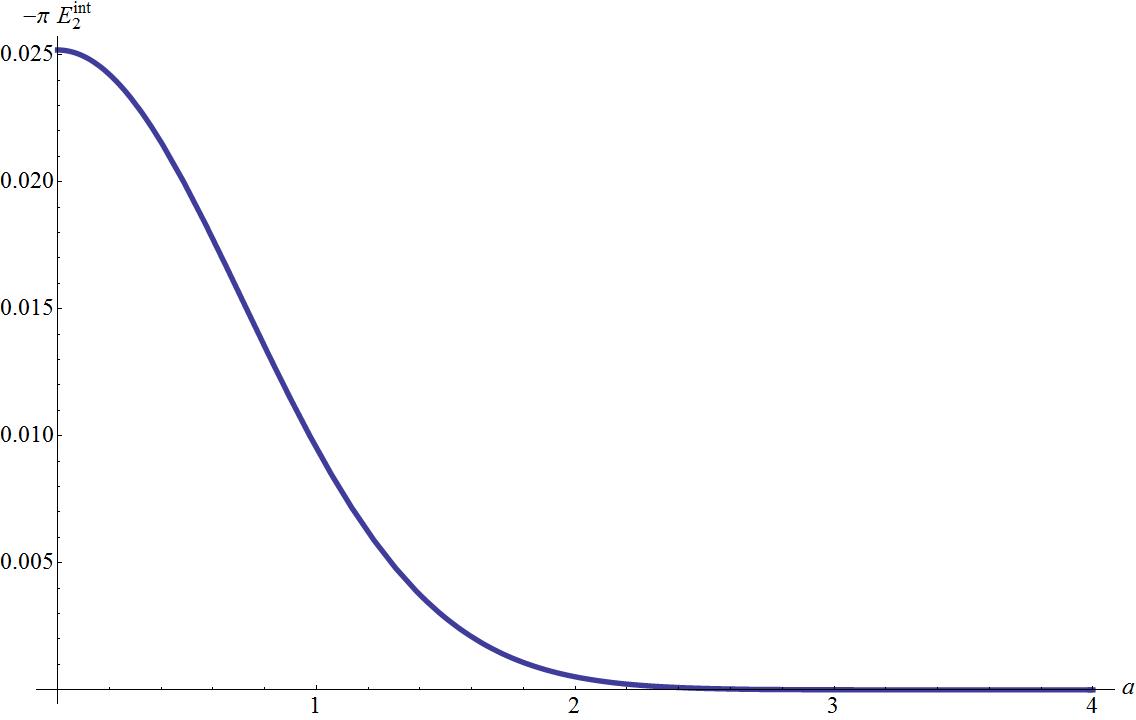}}
\caption{\footnotesize{First order in ${\cal M}_\xi$ (left) and second order in ${\cal M}_\xi$ (right) to the one-loop order interaction energy between two sine-Gordon topological defects.}}
\end{figure}

\begin{figure}[h!]
\centerline{\includegraphics[height=5cm]{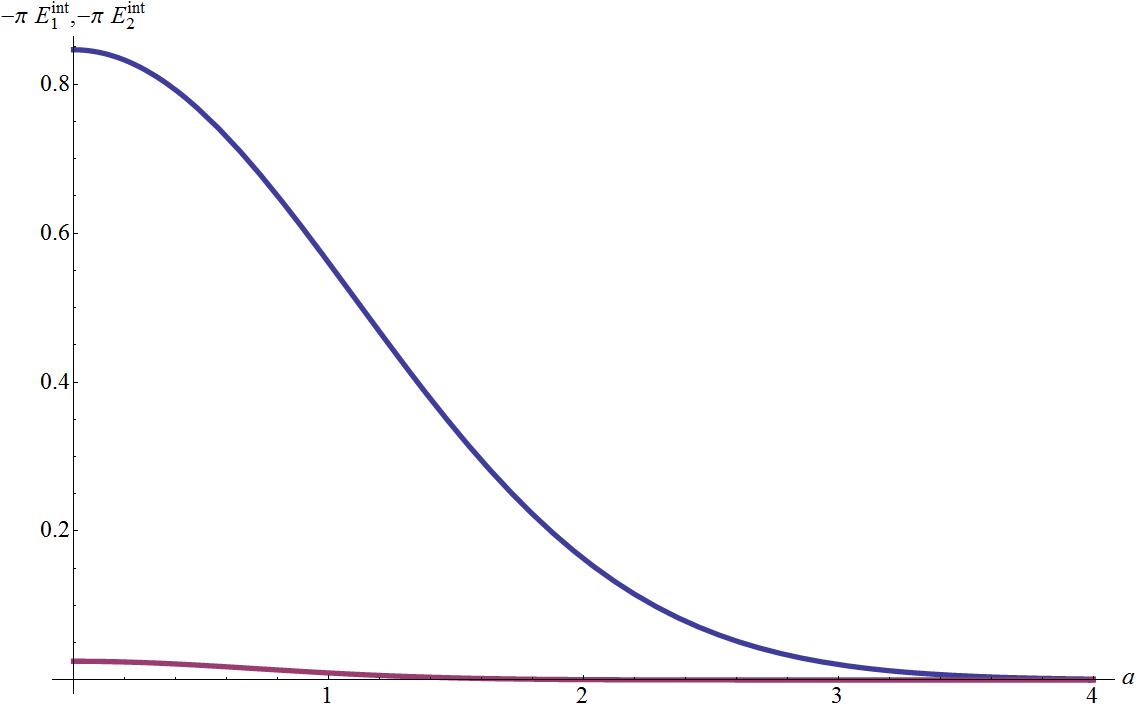}\includegraphics[height=5cm]{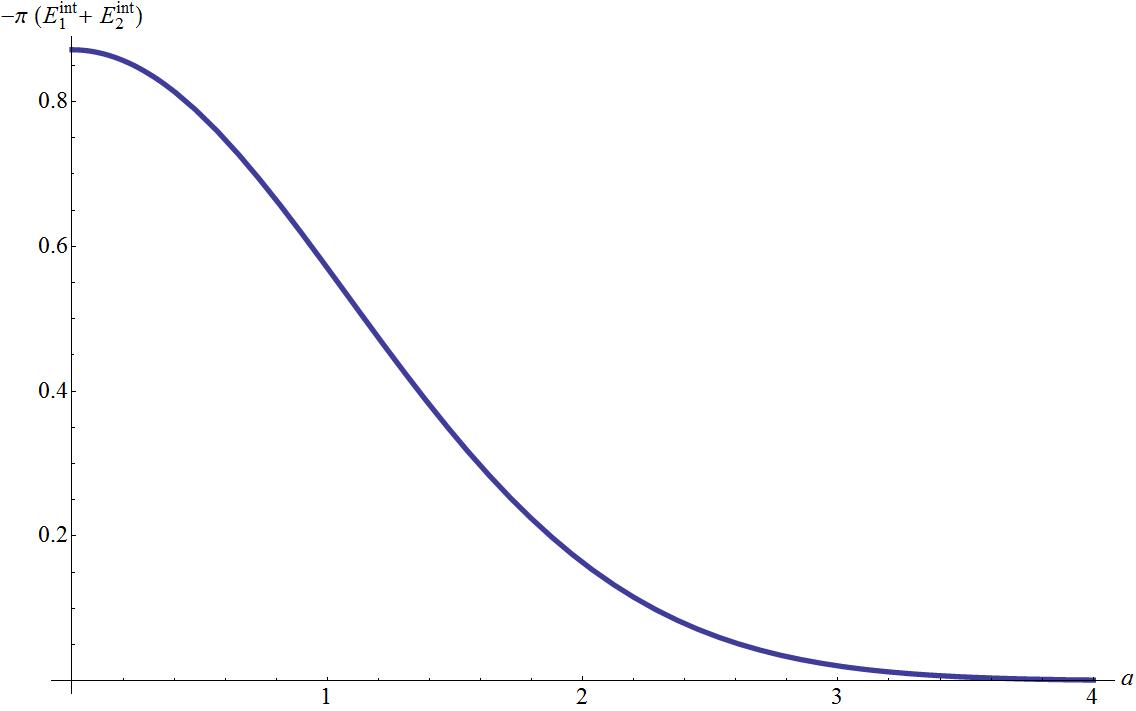}}
\caption{\footnotesize{Comparison between $E^{{\rm int}}_1$ and $E^{{\rm int}}_2$ (left), and one loop correction to the interaction energy between two sine-Gordon topological defects up to second order in ${\cal M}_\xi$ (right).}}\label{fig4}
\end{figure}

\section{Conclusions and outlook}

In the forgoing section we have computed the quantum vacuum interaction of two kinks of the $SG$--equation. We have extended the $TGTG$--method to compute the corresponding interaction energy of non compact objects represented by smooth classical backgrounds. This allowed us to reduce the problem, which initially requires the calculation of the spectrum of the quantum fluctuation in the background of two kinks. That is possible, at least numerically, but constitutes a quite  complicated problem. In opposite, for the $TGTG$--method, it is sufficient to know the quantum corrections in the background of a single kink. For this problem, explicit formulas are available. In this way, the final formula, Eq.  \Ref{4.30}  with \Ref{e1sto}, \Ref{e2ndo} and \Ref{nkink} inserted, involves  3-fold resp. 5-fold integrals over explicit functions. We find an attractive force.

The $TGTG$--formula gives exact results for no-overlapping potentials. In our case, the potentials do overlap. However, since the potentials reach their asymptotic values exponentially fast, the overlap is exponentially small as soon as the separation becomes larger then some times the size of the kinks. Therefor our results are a good approximation for such separations,  for large ones especially.

In opposite to the Casimir effect for conductors, where there is no classical force between the mirrors,   here we have a classical force between the kinks. It can be calculated easily, see Eq. (\ref{3.11}), for instance. For large separation its energy decreases according to
\begin{equation}\label{5.1}
    \Delta_cE_{{\rm int}}(K,K)\sim 32\,e^{-a}.
\end{equation}
For the quantum interaction we observe the same exponential decrease, with a smaller factor in front, however. At all separations, the classical force between two kinks is repulsive, the quantum force is attractive, but small.

Further we mention that the quantum correction is small on the background of the classical part. There are two sources for smallness. The one is the factor in front, see Eq. (\ref{2.7}). The second is the numerical smallness. Here one needs to compare the classical energy shown in Fig. \ref{fig3}, with the quantum one shown in Fig. 4, which is by a numerical factor of approximately 6 smaller. This feature is quite similar to the one with the quantum correction to the mass of the kink. Another remark concerns the expansion of the logarithm in \Ref{4.23}. We calculated the first two orders of its expansion. We have seen that the second order is more than one order smaller than the first one. Since the expansion of the logarithm is equivalent to a perturbative expansion in powers of the background field, we can conclude that such a perturbative expansion works well in the considered example.

We conclude with the remark that the techniques demonstrated in this paper has the potential to simplify calculations of the quantum interaction for more complicated objects like strings and other.

\section*{Acknowledgement}
The authors benefited from exchange of ideas by the ESF Research Network
CASIMIR.\\
This work was supported by DFG grant number BO 1112/18-1.

\bibliography{C:/Users/bordag/WORK/Literatur/bib/papers,C:/Users/bordag/WORK/Literatur/Bordag,C:/Users/bordag/WORK/Literatur/libri,jmmc-libri,jmmc-articoli}
\bibliographystyle{unsrt}
\end{document}